\documentclass[printer]{aa}
\usepackage{txfonts}
\usepackage{graphicx}
\usepackage{amssymb}

\def\aap{{Astronomy \& Astrophysics}}
\def\apj{{Astrophysical Journal}}
\def\apjl{{Astrophysical Journal Letters}}

\def\mnras{{Monthly Notices of the Royal Astronomical Society}}
\def\nat{{Nature}}

\begin{document}

\titlerunning{Tori in Quark-Nova remnants: application to AXPs}
\authorrunning{Ouyed et al.}
\title{Quark-Nova remnants II:\\
 {\small The degenerate torus case with applications to AXPs}}

\author{Rachid Ouyed, Denis Leahy, and Brian Niebergal }

\institute{Department of Physics and Astronomy, University of Calgary, 
2500 University Drive NW, Calgary, Alberta, T2N 1N4 Canada}

\offprints{ouyed@phas.ucalgary.ca}

\date{recieved/accepted}

\abstract{
In the first paper
of this series we explored the case where a quark-nova ejecta forms a degenerate shell, supported by the
star's magnetic field.  Herein, we consider the case where the ejecta has sufficient angular momentum to 
form a torus, and we show that the density and temperature of the torus are such that it will 
remain degenerate throughout it's lifetime.  We go on to discuss the evolution of such a torus
and apply it to AXPs, specifically 1E2259$+$586 and 4U0142$+$615.  As it turns out, using our model we 
can account for many of the observations of these objects including the 
quiescent phase luminosity, and blackbody temperatures during both quiescence and bursting.
Furthermore, for 1E2259$+$586 our model explains the steep and slow decay components seen in the
burst lightcurve, as well as the rotation period glitches and enhanced spin-down rate.
 We also estimate the mass of the degenerate torus to be of the order of
 $10^{-6}M_{\odot}$, and  speculate that the  observed optical/infrared emission from 4U0142$+$615
  might be a signature 
 of the thin degenerate torus  we describe here.
\keywords{accretion, accretion disks -- (stars:) pulsars: general -- dense matter -- X-rays: bursts -- Elementary particles} }

\maketitle

\section{Introduction}

Discussed herein is the novel idea that a torus composed of degenerate matter from a quark-nova
event could be responsible for features of Anomalous X-ray Pulsars (AXPs).
As described in the first paper of this series (Ouyed, Leahy \& Niebergal 2006; hereafter referred to as OLNI),
during a quark-nova the degenerate crust of a neutron star is blown off, leaving behind a quark star (QS)
surrounded by left over, highly-metallic degenerate matter.

In OLNI we discussed one possible fate of this matter;
the case where the ejected crust had insufficient angular momentum to escape the QS's 
gravitational pull. Thus, it would either balance with the QS's magnetic field 
and form a co-rotating shell, or fall back entirely onto the QS.
Using the shell scenario in OLNI, we were able to explain many features of AXPs, Soft Gamma-ray Repeaters (SGRs)
and X-ray Dim Isolated Neutron stars (XDINs).  Also, by appealing to the idea presented in Niebergal et al. (2006), 
where a QS in the ground Color-Flavor Locked (CFL)\footnote{See OLNI on CFL
 and the assumptions in our model.} phase behaves as a type II superconductor and emits X-rays through
magnetic vortex expulsion, a good fit to X-ray luminosities of AXPs and SGRs was made. 


In this second paper the fate of the ejected neutron star crust is considered when the
quark-nova compact remnant has initial conditions (i.e.~rotation period, magnetic field, and shell mass)
such that the propeller mechanism acts.  This, we argue, results in a thin degenerate torus 
forming from the ejected matter rather than a shell as in OLNI.  As it is uncertain what initial conditions 
the quark-nova compact remnant will have, we feel that this second paper represents 
a more complete investigation of the fate of the ejected crust.

We find that although a torus and shell may seem quite similar,
the effects of geometry on bursting and period glitches is significant.
As such, in this paper, we attempt to describe the fracturing of the 
torus due to shear forces caused by differential rotation 
within the torus.  Moreover, we show how this leads to the accretion of the inner edges of the torus
as it is slowly permeated by the star's magnetic field, leading to X-ray bursting.
The nature of this bursting, as seen in the AXP 1E2259$+$586, can be explained within the framework
of our model by appealing to a degenerate torus with a high metallicity, which exhibits changes in it's
ionization fraction as it is bombarded with X-rays during bursts. 
It may be the case this emission, due to the presence of a torus, is insignificant compared to
emission from vortex expulsion. However, we feel that in at least the cases of the two oldest AXPs
(1E2259$+$586 and 4U0142$+$615) the torus emission can exceed emission from vortex expulsion.

This paper is presented as follows: In \S~\ref{sec:torus_formation} we review
 the transport of angular momentum leading to the torus formation and it's evolution. 
In \S~\ref{sec:quiescent_torus} we discuss the quiescent phase due to accretion from the non-degenerate atmosphere
of the torus, followed by the bursting phase in \S~\ref{sec:torus_bursting_phase} due to the star's magnetic field
penetrating the torus's inner edge.
Following that, we apply our model in \S~\ref{sec:cases} to the AXPs 1E2259$+$586 and 4U0142$+$615,
and in \S~\ref{sec:discussion} we provide a brief discussion of the implications and predictions of our model.
  
\section{Propeller regime and torus formation}\label{sec:torus_formation}

As described in Keranen et al. 2005, during the quark-nova the core 
of the parent object converts to (u,d,s) quark matter and becomes 
suddenly compact.  As the core contracts it becomes detached from 
the outer crust.  Consequently the outer crust is left at a radius
larger than the surface of the newly born quark star.  The fate of this
crust depends on the amount of energy released during the quark-nova,
and its coupling to the crust.  

The metal-rich ejecta from the Quark-Nova (QN) is degenerate (described in OLNI),
and remains so as it expands out to the magnetic equilibrium radius, $R_{\rm m}$, which is the 
radius where the ejecta's gravitational pressure balances the star's magnetic pressure.
Once there it will have the form of a shell, and will have expanded only to a thickness of, 
$\Delta R_{\rm m}/R_{\rm m} = 1.2\times 10^{-2}m_{-7}^{1/4}$, 
where the shell's mass in terms of $10^{-7}M_{\odot}$ is $m_{-7}$.  
Provided that the quark star's (QS) magnetic field is strong enough to support this shell, then it will
cool rapidly remaining degenerate, as shown in OLNI.

If this magnetic equilibrium radius is larger than the corotation radius
(which should be possible in at least some instances given the uncertainty in initial conditions
of the quark-nova compact remnant) the propeller mechanism
will take effect (Schwartzman 1970; Illarionov \& Sunyaev 1975),
deflecting the degenerate shell into a torus on the equatorial plane, 
which is the situation considered in this paper. 

Using an angular momentum conservation argument, we can estimate the location
of such a torus by writing 
\begin{equation}\label{eq:omegas}
R_{\rm m}^2 \Omega_{\rm QS, i} = R_{\rm t}^2 \Omega_{\rm K}\ ,
\end{equation}
where $\Omega_{\rm QS, i}$ is the quark star's initial period, 
$R_{\rm t}$ is the equatorial location of the torus in a Keplerian
orbit, and $\Omega_{\rm K} = \sqrt{GM_{\rm QS}/R_{\rm t}^3}$ is the Kepler rotation frequency.
Since all of the shell material is propelled away at the equator with constant 
specific angular momentum leading to a thin torus.
Applying the equation for the magnetic radius (Eq.~1 in OLNI),  
implies a torus radius of,
\begin{equation}\label{eq:dradius}
R_{\rm t} \simeq
 15\ {\rm km}\ \frac{B_{\rm 0, 15}^4  R_{\rm QS,10}^{12}}{ m_{-7}^{2} 
M_{\rm QS, 1.4}^{3} P_{i,ms}^{2}}  \ ,
\end{equation}
where the QS's birth period, $P_{\rm{i,}ms}$, is given in units of milliseconds,
 while the  QS's surface magnetic field strength at birth, $B_{\rm 0, 15}$, is in units
  of  $10^{15}~\rm{G}$.  The radius  and mass are 
$R_{\rm QS,10}$ in units of $10~\rm{km}$, and $M_{\rm QS, 1.4}$ in units of $1.4M_{\odot}$ respectively.
However, in order to form a torus we require enough angular momentum 
transfer to guarantee $R_{\rm t} > R_{\rm m}$. This translates into an upper limit
on the initial period of,
\begin{equation}\label{eq:plimit}
P_{\rm i, ms}  < P_{\rm lim} \sim 2.5\ {\rm ms}
\frac{B_{\rm{0,}15}^{3/2}R_{\rm{QS,}10}^{9/2} }{m_{-7}^{3/4}M_{\rm{QS,}1.4}^{5/4}} \ .
\end{equation}

Moreover, the duration of the propeller phase can be approximated to be 
$t_{\rm prop}\simeq R_{\rm t}/v_{\rm prop}$.
With $v_{\rm prop}\simeq \Omega_{\rm QS}\times R_{\rm m}$ this gives
\begin{equation}
  \label{eq:tprop}
 t_{\rm prop.}\simeq \frac{1}{\Omega_{\rm QS,i}^{1/2}}\frac{1}{\Omega_{\rm K}^{1/2}}\sim 0.15 \ {\rm ms}\
 \frac{P_{\rm i, ms}^{1/2}R_{\rm t,15}^{3/4}}{M_{\rm QS,1.4}^{1/4}}\ .
\end{equation}
 \begin{figure}[t!]
\resizebox{\hsize}{!}{\includegraphics{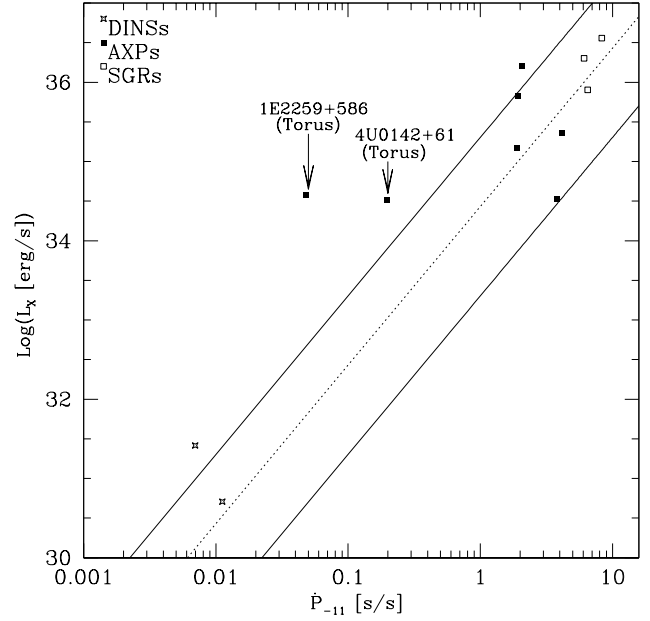}}
\caption{\label{fig:lx}
X-ray luminosity is plotted against period derivative according to Eq.~(30) in OLNI.
The upper and lower solid lines represent a magnetic to X-ray conversion efficiency, 
$\eta_{\rm X}$, of 1 and 0.1 respectively,
and the dashed line is the luminosity averaged over all viewing angles. 
AXP 1E2259$+$586 and AXP 4U0142$+$615 have luminosities
higher than what is  predicted in OLNI from magnetic field decay alone. The higher
luminosity we attribute to steady accretion from a torus.}
\end{figure}

\subsection{Torus fracture from tidal shear}\label{sec:torusfracture}
   
After vertical expansion the torus radial extent (eq.~\ref{eq:drtorus}) is still small.
However, the Keplerian velocity within the torus
varies from $v_{\rm K,1}\simeq \sqrt{GM_{\rm QS}/R_{\rm t}}$ at the equator
to $v_{\rm K,2}\simeq \sqrt{GM_{\rm QS}/1.7R_{\rm t}}$ at $z=\pm R_{\rm t}$ (see eq.~\ref{eq:shape}),
resulting in a large shear. 
 This shear force acts only in the azimuthal direction, or in other words the shear
operates only between surfaces of differing $v_{\rm circ}$.  Thus, the main part of the torus
where density is highest will fracture into subsequently large cylinders, 
or ``walls'' of thickness $\Delta r_{\rm w}$ and mass $m_{\rm w}$. 
Hydrostatic equilibrium along the surfaces of the walls guarantees that each wall will remain
 in high pressure contact with the next. In other words we do not
 expect a discontinuous set of walls but a rather walls separated by melted (non-crystalline) material,
or basically a degenerate fluid.
 
To determine the thickness of these walls, one appeals to 
the ultimate strength of iron, which is $U = \psi E_{\rm B}$, where $E_{\rm B}\sim 1$ MeV is the
binding energy per nucleon of the torus material, and $\psi$ is the strain factor 
$\psi\sim 10^{-3}$ for Iron (e.g. Halliday \& Resnick  \S 13.6).
By equating the work performed by the shear forces $W = F\times \Delta r_{\rm w}$,
(where the tidal force per unit area  is $F = - 2 GM_{\rm QS}  \Delta r_{\rm c} / r_{\rm w}^3$) with the
torus binding energy, $E_{\rm B, MeV} m_{\rm w}/(56 m_{\rm H})$,
one finds a  typical wall thickness to be 
$\Delta r_{\rm w} \sim 400\ {\rm cm}\  f_{\rm Fe} r_{\rm w,15}^{3/2}$.   
Here $f_{\rm Fe}= \psi_{-3}^{1/2} E_{\rm B, MeV}^{1/2} /M_{\rm QS,1.4}^{1/2}$, and
$r_{\rm w,15}$ is the distance of the wall from the center of the quark star in units of $15~\rm{km}$.
Thus, as a fiducial example, the innermost wall ($r_{\rm w}=R_{\rm t}$) has a thickness of,
\begin{equation}\label{eq:drring}
\Delta r_{\rm w} \sim 400\ {\rm cm}\  f_{\rm Fe} R_{\rm t,15}^{3/2}\ .
\end{equation}

The number of walls in the entire torus is then dependent on the thickness of the torus, 
and can be found to be,
\begin{equation}\label{eq:nwalls}
N_{\rm w} \sim 2000 \frac{T_{\rm keV}^{5/4}}{f_{\rm Fe} R_{\rm t,15}^{3/2}} \ .
\end{equation}
Also, the mass of the innermost wall in terms of the torus equilibrium temperature
(equation \ref{eq:Teq}) is,
\begin{equation}\label{eq:mwall}
 m_{\rm w} \simeq    6.5\times 10^{-11} M_{\odot} \frac{\mu_{\rm atm}^{15/8}f_{\rm Fe} m_{-7} R_{\rm t,15}^{1/4}}{\eta_{0.1}^{5/4} t_{\rm yrs}^{1/2}}  \ ,
\end{equation}
 where $\mu_{\rm atm}$ is the mean molecular weight of the torus atmosphere discussed in Appendix A
  and 
in the following section.
The time dependency in the  equation above is due to the average density of the torus
decreasing in time as the torus spreads radially in a manner given by Eq.~(\ref{eq:drtorus}). 
There is also a small extra dependency in time due to the torus mass gradually decreasing
from torus's atmosphere being accreted during the quiescent phase, thus reducing the total torus mass
as given by equation (\ref{eq:mdot_q}).

\section{The quiescent phase}\label{sec:quiescent_torus}

As explained in Niebergal et al. (2005), the 
 X-ray luminosity during the quiescent phase in our model
is due to vortex expulsion.  The magnetic field contained within
the vortices is also expelled, and the subsequent magnetic reconnection leads
to the production of X-rays with a luminosity as given in OLNI, 
\begin{equation}\label{eq:lx}
L_{\rm X} \simeq 2.01\times 10^{35}\ {\rm erg\ s}^{-1} \eta_{\rm X} \dot{P}_{-11}^2 \ ,
\end{equation} 
where $\eta_{\rm X}$ is the efficiency parameter inherent in the coversion from magnetic
energy to radiation.   Figure 1 (reproduced from  OLNI) shows our 
model of luminosty evolution as compared to the quiescent X-ray 
luminosities of AXPs, SGRs and XDINSs. It is clear that 
 AXP 1E2259 and AXP 4U0142 show an excess luminosity as compared to
SGRs/AXPs born with a shell. This we attribute to continuous accretion
from the torus as explained below.

\subsection{Steady accretion from the torus}

At the edges of the torus, using equation (\ref{eq:rhotorus}), the density is low enough such that the torus material becomes non-degenerate,
which creates a thin metal-rich atmosphere.  A fraction of this atmosphere is constantly accreted 
onto the QS, where it is converted into CFL quark matter, releasing the excess energy as radiation.
This radiation in turn heats up the torus, thus altering the size of the non-degenerate atmosphere.

For a given temperature, material from the non-degenerate torus atmosphere 
slowly leaks out at a rate, $\dot{m}$. 
For any non-zero $\dot{m}$, the accretion luminosity heats up the torus, 
which also cools rapidly as a blackbody.
Therefore, one can determine the equilibrium temperature given by Eq.~(\ref{eq:Teq}), 
\begin{equation}\label{eq:Teq2}
T_{\rm eq} \sim  0.52 \ {\rm keV} \frac{\eta_{0.1}R_{t,15}}{\mu_{\rm atm}^{3/2}M_{\rm QS,1.4}}\ .
\end{equation}
This also implies a continuous or equilibrium accretion rate
from the torus atmosphere of,
\begin{equation}\label{eq:mdot_q}
  \dot{m}_{\rm eq}\sim  2.3\times 10^{16}\ {\rm gm\ s}^{-1} 
  \frac{\eta_{0.1}^{3}R_{\rm t, 15}^{6}}{\mu_{\rm atm}^{6}M_{\rm QS,1.4}^{4}} \ .
\end{equation}
Thus, the corresponding equilibrium accretion luminosity is,
\begin{equation}\label{eq:lxnd}
 L_{\rm acc} =  \eta \dot{m}_{\rm eq}c ^2 \sim  2.2\times 10^{36}\ {\rm erg\ s}^{-1} 
 \frac{\eta_{0.1}^{4}R_{\rm t, 15}^{6}}{\mu_{\rm atm}^{6}M_{\rm QS,1.4}^{4}}\ .
\end{equation}

This defines the quiescent phase luminosity in our model for a QS with a torus. 
From this, a blackbody temperature from the accretion hot spot on the star's surface can be found to be,
\begin{equation}\label{eq:temp_bb}
 T_{\rm BB,eq} \sim  0.92\ {\rm keV}
 \frac{\eta_{0.1} R_{\rm t,15}^{3/2}}{\mu_{\rm atm}^{3/2}R_{\rm BB,5}^{1/2}M_{\rm QS,1.4}}\ ,
\end{equation}
where $R_{\rm BB,5}$ is the accretion region (i.e. the polar cap since most of the accreted material
is channeled towards the poles) in units of 5 km.

\subsection{The effect of mean molecular weight, $\mu$}\label{sec:effect_of_mu}


Ionization of the torus atmosphere is due to  X-ray photons from 
accretion onto the poles as mentioned above.  
Iron ionization equilibrium has been discussed at length in Arnaud \& Raymond (1992),
who show (see their figure 10) the most abundant charge states versus temperature of iron.
For the equilibrium temperatures in the torus (Eq.~\ref{eq:Teq2}), 
we estimate that the charge state is situated between the
L and M shells, which corresponds to $\mu_{\rm Fe}$ values of $\sim 3.3$.
 For comparison fully ionized iron yields  $\mu_{\rm Fe}\sim 2.07$.

Therefore during the quiescent phase ($\mu\sim 3.3$), our model predicts the following parameters and observables,
\begin{eqnarray}\label{eq:lxnds}
  \dot{m}_{\rm eq}&\sim&  1.9\times 10^{13}\ {\rm gm\ s}^{-1} 
  \frac{\eta_{0.1}^{3}R_{\rm t, 15}^{6}}{M_{\rm QS,1.4}^{4}}\\\nonumber
 T_{\rm eq} &\sim& 0.085\ {\rm keV}\ \frac{\eta_{0.1} R_{\rm t,15}}{M_{\rm QS,1.4}} \\\nonumber
 T_{\rm BB,eq} &\sim&  0.15\ {\rm keV}
 \frac{\eta_{0.1} R_{t, 15}^{3/2}}{R_{\rm BB,5}^{1/2}M_{\rm QS,1.4}}\\\nonumber 
  L_{\rm acc} & \sim&   1.7\times 10^{33}\ {\rm erg\ s}^{-1} 
  \frac{\eta_{0.1}^{4}R_{\rm t, 15}^{6}}{M_{\rm QS,1.4}^{4}}\ .
 \end{eqnarray}

Clearly, in the case where a torus is formed around a QS, as opposed to the shell
case presented in OLNI, the quiescent phase is dominated by continuous accretion luminosity,
not by vortex expulsion.  The observed quiescent luminosities of AXP 4U0142 
($\simeq 3.3\times 10^{34}\ {\rm erg\ s}^{-1}$) and AXP 1E2259 
($\simeq 3.8\times 10^{34}\ {\rm erg\ s}^{-1}$; see Table I in OLNI) 
implies the location of these tori to be at roughly $R_{\rm t}\sim  25$ km. 
These two candidates are discussed in more details in \S~\ref{sec:cases}.
 

In principle, for a given source (i.e. fixed $R_{\rm t}$) a change
 by a factor of a few  in the mean molecular weight of the torus
  atmosphere 
  implies large changes in the accretion luminosity from the torus since
  \begin{eqnarray}
  T_{\rm t, eq} &\propto& \mu_{\rm atm.}^{-3/2}\\\nonumber
  T_{\rm BB, eq} &\propto& \mu_{\rm atm.}^{-3/2}\\\nonumber
 L_{\rm t, acc.} &\propto& \mu_{\rm atm.}^{-6}\ .
  \end{eqnarray}
  For example if the degenerate atmosphere is suddenly heated
   as to become fully ionized  during a bursting
   episode (discussed in the next section), then $\mu_{\rm atm}$ would
   drop from 3.3 to 2.1  yielding,
     \begin{eqnarray}
  \frac{T_{\rm b}}{T_{\rm eq}} \sim 2.0\\\nonumber
  \frac{T_{\rm BB, b}}{T_{\rm BB, eq}} \sim 2.0\\\nonumber
\frac{L_{\rm b, acc.}}{L_{\rm acc.}} \sim 16.4\ ,
  \end{eqnarray}
where the subscript ``b" stands for bursting.
   During the bursting phase then, the torus equilibrium temperature  and the 
     blackbody temperature of the hot spot on the star 
  would have increased by a factor $\sim 2$ while the 
  torus accretion luminosity would  increase by  over an order of magnitude
  from its quiescent value. The mean molecular weight can also change
   due to a change in composition leading to enhanced temperature
    and luminosity (see \S \ref{sec:4.4}).

\section{The bursting phase}\label{sec:torus_bursting_phase}

In our model, the bursting phase is initiated when the innermost wall of the torus 
is magnetically permeated such that it detaches from the torus itself, 
where it is then accreted rapidly onto the QS.  

\subsection{Magnetic field penetration}
   
Recall from \S~\ref{sec:torusfracture} that the torus is composed of subsequently larger walls,
made of a solid degenerate matter, and separated by a degenerate fluid.
As such, the torus's inner wall will be slowly penetrated by the QS's magnetic field on 
timescales determined by the induction equation,
\begin{equation}
   \frac{\partial B}{\partial t}  =  \frac{c^2}{4\pi \sigma} \nabla^2 B\ .
\end{equation}
Here, $\sigma = n_{\rm e, th} e^2 \lambda_{\rm e}/(m_{\rm e} v_{\rm rms})$
and $\lambda_{\rm e}=1/(n_{\rm e, th}\sigma_{\rm T})$, where $n_{\rm e, th}$ 
is the number density of thermal electrons in the degenerate matter, 
$\sigma_{\rm T}$ is the Thompson scattering cross-section, 
and the root-mean-square electron velocity is $v_{\rm rms}= c_{\rm s}$. 
Therefore, the time needed for the magnetic field to penetrate to a depth of $\Delta r_{\rm w}$ into the torus is, 
\begin{equation}\label{eq:btime}
 \tau_{\rm B} \sim 82 \ {\rm yrs}\ \frac{f_{\rm Fe}^{2}  R_{\rm t,15}^{3}}{\rho_{\rm t,7}^{1/6}} \ ,
\end{equation}
where, $\Delta r_{\rm w}$, is the wall's thickness (see Eq.~\ref{eq:drring}), and equation~\ref{eq:rhot} was used.
As the torus spreads in time its density decreases as $t^{-1/2}$ (from Eq.~\ref{eq:rhot}), thus weakly  
decreasing the penetration timescale as the source ages by, $\tau_{\rm B}\propto t^{-1/12}$.

\subsection{Accretion of the wall}\label{sec:wallaccretion}

Once the magnetic field penetrates the innermost wall, the poloidal component
of the dipole field is wound up introducing an azimuthal component $B_{\phi}$.
 The amplification of the magnetic field can be found from $\partial {\bf B} /\partial t
 = \nabla \times ({\bf v}\times {\bf B})$ which yields $\partial B_{\phi}/\partial t\sim \Omega_{\rm K} B_{\rm p}$,
  where $\Omega_{\rm K}=\sqrt{(GM_{\rm QS}/R_{\rm t})}$ is the Kepler frequency and $B_{\rm p}$ 
is the poloidal component of the QS's dipole field.
  That is,  the magnetic field within the wall will be amplified following the relation 
$B_{\phi} \propto \Omega_{\rm K} B_{\rm p} t$,  until the magnetic energy is comparable to the kinetic energy
of the wall,
at which point it dominates the dynamics. This rapid build up of $B_{\phi}$ occurs roughly on the timescale
of milliseconds.

The resulting  magnetic torque  ($\propto R_{\rm t}^{2} B_{\phi}B_{\rm p}$) spins-down, the wall 
and spins-up the star. Or in other words, the magnetic torque is essentially separating 
the wall from the main body of the torus and transferring it's angular momentum
to the star, inducing a rotation period anomaly (called a ``glitch''; 
discussed further in \S~\ref{sec:period_changes}).  This causes the wall to eventually co-rotate with the QS.

The  wall's initial mass at the time of detachment from the torus, $m_{\rm w,i}$,  
is given by equation (\ref{eq:mwall}) (using $\mu_{\rm atm}=3.3$),
\begin{equation}\label{eq:init_wall_mass}
m_{\rm w,i} \simeq    6.1\times 10^{-10} M_{\odot} 
\frac{f_{\rm Fe} m_{\rm -7} R_{\rm t,15}^{1/4}}{\eta_{0.1}^{5/4} t_{\rm yrs}^{1/2}}  \ .
\end{equation}
Thus, as the the gravitational pressure of the wall is much weaker than the magnetic pressure from the QS,
the co-rotating wall is trapped between the torus and the pressure of the underlying 
magnetic field.  Furthermore, the effective gravity acting on the
wall's edges also guarantees that the wall is pressure confined vertically. Moreover, by 
equating the magnetic pressure to the wall's pressure, $B^2/(8\pi)= \kappa {\rho_{\rm w}^{c}}^{4/3}$
where $\kappa= 1.24\times 10^{15} \mu_{\rm e}^{-4/3}$, we find the density of the wall after detachment to be,
\begin{equation}
 \rho_{\rm w}\sim 1.38\times 10^{8}\ {\rm gm\ cm}^{-3}\ 
\frac{B_{\rm s,14}^{3/2} R_{\rm QS,10}^{9/2}}{R_{\rm t, 15}^{9/2}}\ ,
\end{equation}
where the subscript ``w" implies a co-rotating wall and depicts the fact that the detached
wall is spun-down to co-rotate with the star.  The value of the magnetic field in equation
above is evaluated at detachment and is given in our model as 
$B_{\rm s}=\sqrt{3\kappa_{\rm B}P\dot{P}}$ where $\kappa_{\rm B}= 8.8\times 10^{38}\ {\rm G}^{2}\ {\rm s}^{-1}$ 
 (see Niebergal et al. 2006; and also OLNI).
 
Given this situation, the wall will lose mass by 
thermal evaporation  of it's non-degenerate atmosphere, thus getting thinner in time while
keeping the same density and pressure (i.e.~$\Delta r_{\rm w} = 2\pi R_{\rm t}^2 m_{\rm w}(t)/\rho_{\rm w}$).
By appealing to equations (\ref{eq:mdotwall}) \& (\ref{eq:Twall}) in the appendix 
we arrive at the following equation for the decrease  in mass of the wall 
(expressed here in units of $10^{-10}M_{\odot}$) over time,  
\begin{equation}
\frac{d m_{\rm w,-10}}{{(m_{\rm w,-10}})^{2}} = 
- 7.7\times 10^{-4} \frac{\eta_{0.1} R_{\rm t,15}^{5}}{\mu_{\rm w, atm} B_{\rm s,14}^{3} R_{\rm QS,10}^{9}} dt\ .
\end{equation}
Thus, the wall is  accreted in time at a rate of,
\begin{equation}\label{eq:mctime}
\frac{m_{\rm w,i}}{m_{\rm w}} = 1 + \frac{t}{\tau_{\rm w}}\ ,
\end{equation}
where a characteristic time for wall consumption is defined to be,
\begin{equation}\label{eq:wallt}
\tau_{\rm w} = 0.4\ {\rm hours}\ \frac{\mu_{\rm w, atm} B_{\rm s,14}^{3} 
R_{\rm QS,10}^{9}}{\eta_{0.1} R_{\rm t,15}^{5} } \left(\frac{10^{-10}M_{\odot}}{m_{\rm w,i}}\right)\ .
\end{equation}
 The expression for $\tau_{\rm w}$ implies that the more massive the wall at detachment the faster
it gets consumed. Equation (\ref{eq:mctime}) shows that 99\% of the wall
is consumed in $ 100\tau_{\rm w} $. 
 The accretion of the wall gives a much higher accretion rate (see eq.(B1)) 
 compared to the quiescent phase accretion from the torus atmosphere. The
 resulting equilibrium temperature  (eq.(B2))  is high enough
  to fully ionize the wall atmosphere, resulting in a $\mu_{\rm w,atm}\sim 2.1$
  while the wall is being consumed.

\subsection{The first burst component: the steep decay}

The first component in our model is defined by consumption of  the detached wall.
 Using equation (\ref{eq:mctime}), the corresponding luminosity due to accretion 
of the wall from its atmosphere varies in time as,
\begin{equation}\label{eq:lumin_fast_decay}
  L_{\rm w}\left(t\right) = \eta \dot{m}_{\rm w} c^2 = \frac{L_{\rm w,0}}{(1+\frac{t}{\tau_{\rm w}})^2}\ ,
\end{equation}
where the initial luminosity due to the accretion of the wall is,
\begin{equation}\label{eq:wall0_lumin}
  L_{\rm w,0} \sim 5\times 10^{39}\ {\rm erg\ s}^{-1}\
  \frac{\eta_{0.1}}{\tau_{\rm w, hrs}} \left(\frac{m_{\rm w,i}}{10^{-10}M_{\odot}}\right) \ ,
\end{equation}
and $\tau_{\rm w, hrs}$, is expressed in hours. The wall
temperature as given by equation (B2) starts at a few keV and
drops as $m_{\rm w}(t)^{1/2}$ until it becomes negligible
compared to the steady emission due to the torus accretion. See Figure (\ref{fig:2comp}) 
for a comparison with observations.

During the entire accretion process the amount of energy released will be, 
\begin{eqnarray}\label{eq:echunk}
E_{\rm w} &=& \int^{\infty}_{0} L_{\rm w}\left(t\right) dt = \tau L_{\rm w,0} \\\nonumber
 &\sim&  5\times 10^{39}\ {\rm erg}\ \eta_{0.1} \left(\frac{m_{\rm w,i}}{10^{-10}M_{\odot}} \right)  \ .
\end{eqnarray}

\subsection{The second burst component: the slow decay}
\label{sec:4.4}

   During the burst, the  torus is irradiated by MeV neutrinos
    and photons causing the dissociation of the iron in the torus.
    For a typical total energy of about $10^{39}$ erg ($\sim  10^{45}$ MeV)
 released during the consumption of the detached wall (see Eq. \ref{eq:echunk}), we estimate
  about 1 iron nucleus dissociated per  MeV, or a total of
  $\sim  10^{45}$ dissociations. Hydrostatic buoyancy then 
   causes the nuclei with $Z\sim26/2= 13$ to float to the top 
of the torus onto the non-degenerate atmosphere. 
 Following  the wall evaporation and torus irradiation, the torus atmosphere is now
  mostly  composed of nuclei with $Z\sim13$ instead 
 of $Z=26$. This result in decreased $\mu_{\rm atm}$ since the lighter nuclei are fully
 ionized (they have lower 
atomic energy levels by a factor of $Z^2\sim 4$) causing
$\mu_{\rm atm}$ to drop  as low as $\mu_{\rm atm,b}\sim 2.1$  (in the
extreme case of a pure light  nuclei atmosphere) from $\mu_{\rm atm,q}\sim 3.3$.

Over time, $\mu_{\rm atm}$ increases again to 3.3 as the $ 10^{45}$ light nuclei in the atmosphere are 
 slowly depleted by accretion while the 
 atmosphere gradually becomes enriched  to iron again.
 A simple model for light nuclei depletion yields $dX_{13}/dt  = - X_{13}/\tau_{13}$
  where $X_{13}$ is the fractional abundance by mass of the light nuclei in the atmosphere
   and $\tau_{13}$ is the exponential decay timescale.
 This  timescale  is the mass in light nuclei, $m_{\rm 13}\sim  3\times 10^{46}  m_{\rm H}$,  
 in the atmosphere divided by the depletion rate given by $\dot{m}_{\rm t}$, or, 
\begin{equation}
\label{eq:tau13}
\tau_{\rm 13} = \frac{m_{\rm 13}}{\dot{m}_{\rm t}} \sim 13 \ {\rm year} 
\frac{M_{\rm QS,1.4}^4}{\eta_{0.1}^3 R_{\rm t,15}^6 } \ ,
\end{equation}
found using an average of $2.5$ for the mean molecular weight per electron.

 We model the mean molecular weight using two species: partially ionized
 iron ($A=56, Z=26, Z_{\rm e}=16$), and fully ionized light nuclei ($A=28,Z=Z_{\rm e}=13$) where $Z_{\rm e}$
  is the number of free electrons per nucleus of charge $Z$. The corresponding molecular
  weight of the mixture is $\mu^{-1} = \sum_{i} X_{i} (1+Z_{e, i})/A_{i}$ which after substitution
   for $X_{13}(t)$ and $X_{26} =1-X_{13}$ yields
\begin{equation}\label{eq:mu_inv}
  \frac{1}{\mu_{\rm atm} (t)} = \frac{1}{\mu_{\rm atm,q}}  + \left(\frac{1}{\mu_{\rm atm,b}}-\frac{1}{\mu_{\rm atm,q}}\right)  e^{-t/\tau_{13}}\ .
\end{equation}
 With the use of equation (\ref{eq:lxnd}), the luminosity due to accretion during a burst is then,
\begin{equation}\label{eq:second_comp}
L_{\rm b,acc} \sim  L_{\rm q,acc} \left(1+ (\frac{\mu_{\rm atm,q}}{\mu_{\rm atm,b}}-1)e^{-t/\tau_{13}}\right)^{6} \ ,
\end{equation}
 where $L_{\rm q,acc}$ is the accretion luminosity during the quiescent phase
   when the  torus mean molecular weight is $\mu_{\rm atm,q}\sim 3.3$ (see
  Eq. \ref{eq:lxnd}).  A comparison with observations using this is shown in
Figure (\ref{fig:2comp}).



\section{Changes in Rotation Period}\label{sec:period_changes}

\subsection{Instantaneous spin-up during a burst}
\label{sec:5-1}
  
After the star's magnetic field has permeated the innermost wall of the torus,
the wall will detach and begin to co-rotate with the star.  During this transition,
by conservation of angular momentum, the star's period will decrease by an amount given
by the expression,
\begin{equation}
 \frac{\Delta P}{P} = \frac{\Delta I}{I} - \frac{\Delta L}{L} 
 \approx -\frac{5}{2}\frac{ m_{\rm{w}}}{M_{\rm QS}}
   \frac{R_{\rm t}^2}{R_{\rm QS}^2}\frac{\Omega_{\rm t}}{\Omega_{\rm QS}}\ ,
\end{equation}
which can be instead expressed in terms of the star's rotation period (in units of 10 seconds) by,
\begin{equation}\label{eq:deltap}
  \frac{\Delta P}{P} \simeq  -3.8\times 10^{-6} 
  \frac{P_{10}R_{\rm t,15}^{2}}{M_{\rm QS,1.4}R_{\rm QS,10}^{2}}
  \left(\frac{m_{\rm w,i}}{10^{-10}M_{\odot}}\right) \ .
\end{equation}

It is worth pointing out that the period glitch is not necessarily 
simultaneous with the burst, since there may be a period of pressure
adjustment before the wall begins being consumed. 
Thus, a clear prediction is that QSs born with a
torus instead of a shell will show glitch activity just prior to their bursts.
  
\subsection{Persistent spin-down following bursting}

A portion of the non-degenerate wall atmosphere matter is kept in co-rotation by the
magnetic field out to the light cylinder, which provides an efficient mechanism for
removing angular momentum from the system, via the propeller.
The angular momentum per unit mass lost at the light cylinder is $c^2/\Omega_{\rm QS}$,
which enhances the spin-down rate of the quark star during bursts to,
\begin{equation}\label{eq:pdot_persist}
 \dot{P}_{-13,{\rm b}} \simeq 2 \dot{m}_{w,10}  P_{\rm 10}^3 \ . 
\end{equation}
Here, the wall's evaporation or mass-loss rate, $\dot{m}_{w,10}$, 
is given in units of $10^{10}\ {\rm g\ s}^{-1}$, and the enhanced spin-down
rate is in units of $10^{-13}~\rm{s~s}^{-1}$.

\section{SGRs and AXPs in our model:  Case study}\label{sec:cases}
 
Here we specifically focus on two AXPs, namely 1E2259$+$586 and 4U0142$+$615,
(see Table 1 in OLNI for more on their observed features). These two candidates, 
as can be seen in Figure 1, show luminosity during their quiescent phases 
above the luminosity predicted by vortex expulsion.  In our model
this excess luminosity is due to the presence of a torus, which is constantly being
accreted.  Evidence for the presence of a torus has indeed already been detected
around AXP 4U0142$+$615 (Hulleman et al. 2000; Wang et al. 2006).
  
\subsection{AXP4U0142$+$615}

Observations of pulsed optical emission from AXP 4U0142$+$61 has led to
the application of the irradiated disk model, wherein optical and infrared
luminosities are reprocessed emission from an X-ray irradiated disk
(i.e.~Hulleman et al. 2000a).
Most irradiated disk models predict an optical emission larger than what is observed 
in 4U0142$+$61 (Hulleman et al. 2004), so to compensate one either assumes a
larger inner disk radius or a smaller outer radius.  The former implies the
absence of a hot inner region which is inconsistent with observations, 
leaving the option of an outer disk radius very close to the inner radius.   
Thus, one has the possibility of a very thin passive disk, which we argue
fits all of the features of the degenerate torus as described in this paper. 
Normally a very thin disk is considered unrealistic due to heating by viscous dissipation
(Hulleman et al. 2000b), however, in our model 
viscous dissipation is negligible due to degeneracy in the disk.
 
Moreover, if the blackbody component is to be believed, then the emission region is 
confined to a radius of 12 km (White et al. 1996; assuming a distance of 5 kpc),
ruling out some magnetospheric emission models.  In our model, an emission region of this size
is predicted, and confines parameter values that fit well with other observations (see Eq.~\ref{eq:temp_bb}).

Using the observed quiescent phase luminosity of AXP 4U0142$+$61 
($L_{\rm acc} = 3.3\times 10^{34}~\rm{erg~s}^{-1}$), we find $R_{\rm t}\simeq 25$ km.  From this, our model yields
 the equilibrium temperature of the torus, 
blackbody temperature of the emitted X-ray luminosity,  time in years between bursts, and the age of the
system:
\begin{eqnarray}
\nonumber R_{\rm t} &\simeq & 25~\rm{km}\\
\nonumber T_{\rm eq} &\simeq& 0.14~\rm{keV} \\
\nonumber T_{\rm BB,eq} &\simeq& 0.42~\rm{keV} \\
 \tau_{\rm B}&\simeq& 150\ {\rm years}\\
 t_{\rm age} &\simeq& 4.7\times 10^{4}\ {\rm years}
\end{eqnarray}
A $5$ km blackbody emitting radius was assumed to get the blackbody temperature.
 When a burst occurs in the future, equation (\ref{eq:echunk}) gives us
 the wall mass then equation (\ref{eq:deltap})  yields 
  the  expected glitch, $\Delta P/P$, in our model. Using  $R_{\rm t}\simeq 25$ km 
   we find   $\Delta P/P \sim - 9\times 10^{-6} (m_{\rm w, i}/10^{-10}M_{\odot})$.
    The torus mass  can then be derived from equation (\ref{eq:init_wall_mass}) to be
    $m_{\rm t} \simeq 2\times 10^{-6}M_{\odot} (m_{\rm w, i}/10^{-10}M_{\odot})$ which
     is of the order of $10^{-6}M_{\odot}$ for our fiducial wall mass.

\subsection{AXP1E2259$+$586}

During the June 2002 outburst (Woods et al. 2004), AXP 1E2259$+$586 displayed over 80 X-ray bursts
for approximately 4 hours with bursts ranging in duration from 2 ms to 3 s. 
 We refer to this as the first component which  decayed  approximately as 
 a power law in time, $\propto t^{-4.8}$. Enhanced flux was observed over
the following year, referred to as the second component, which decayed 
according to a power law in time $\propto t^{-0.22}$.  
The X-ray properties of the bursts are very similar to those seen in Soft Gamma Repeaters (SGRs).
 This is natural  in  our model since SGRs and AXPs involve the same underlying engine,
namely a quark star accreting from a degenerate shell/torus. We thus propose that the only difference between SGRs and the two AXPs mentioned in this paper,
is that the SGRs were born as QSs with a slower rotation period, therefore possessing a co-rotating shell
(as discussed in OLNI), rather than a thin  torus. 

Observations  give the quiescent phase blackbody temperature 
 of about 0.42 keV. Given the observed quiescent phase luminosity our model implies
 $R_{\rm t}\simeq 25$ km. From this, our model yields
 the equilibrium temperature of the torus, 
blackbody temperature of the emitted X-ray luminosity, 
 time in years between bursts, and the age of the system: 
\begin{eqnarray}
\nonumber R_{\rm t} &\simeq & 25~\rm{km}\\
\nonumber T_{\rm eq} &\simeq& 0.14~\rm{keV} \\
\nonumber T_{\rm BB,eq} &\simeq& 0.42~\rm{keV} \\
 \tau_{\rm B}&\simeq& 100\ {\rm years}\\
  t_{\rm age} &\simeq& 1.5\times 10^{5}\ {\rm years}
\end{eqnarray}

  During the bursting phase and at the onset of the outburst 
the measured temperature increased to $1.7$ keV, before decaying to 0.5 keV within the first few days. 
 This is associated with the wall accretion episode in our model.
Gavriil et al. (2002) reported that they  simultaneously observed increases of the 
pulsed and persistent X-ray emission by over an order of magnitude relative to quiescent levels. 
This increase is also consistent with our model because, as discussed in \S~\ref{sec:effect_of_mu}, 
 reduction in the mean molecular weight of the torus
 atmosphere during the burst causes a significant increase
in the  temperature ($T_{\rm BB}\propto \mu^{-3/2}$ ) and luminosity (i.e. $L\propto \mu^{-6}$).

Further observations of AXP 1E2259$+$586 show that it underwent a sudden spin-up ($\Delta P/P = -4 \times 10^{-6}$), 
followed by a factor of 2 increase in spin-down rate, which persisted for more than 18 days. 
 The spin-up is consistent with the torque due to wall detachment from the torus (see \S \ref{sec:5-1}).
Using equation (\ref{eq:deltap}) and the measured period glitch, one gets for the mass of the detached wall,
$m_{\rm w}\simeq 0.5\times 10^{-10}M_{\odot}$, which is in good agreement 
  with the  expected wall mass based on shear forces on the torus
(Eq.~\ref{eq:init_wall_mass}).
Moreover, using this wall mass, our model predicts (using Eq.~\ref{eq:echunk}) 
the energy in the burst to be $2\times 10^{39}~\rm{erg}$, which is an order of magnitude greater than the
observed burst, implying that observations likely missed the early phases of the burst.
 In addition the age and the wall mass allow us to calculate the torus
 mass from equation (\ref{eq:init_wall_mass}) to be of the order\footnote{Interestingly, Wang et al. (2006)
 estimate the mass of the disk surrounding 4U0142$+$615  to be within the same
 range as the torus mass in our model.  We are tempted to speculate that 
  the discovered debris disk is in fact the degenerate torus described
  in our model. Planet formation around such a metal-rich degenerate torus
  has been suggested and discussed in Ker\"anen\&Ouyed (2003).}
 of $m_{\rm t}\sim 2.2\times 10^{-6}M_{\odot}$.
  \begin{figure}[t!]
\resizebox{\hsize}{!}{\includegraphics{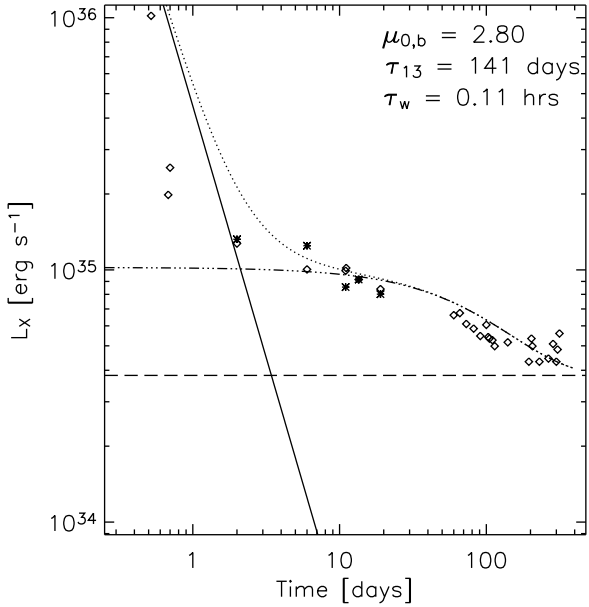}}
\caption{\label{fig:2comp}
Shown is the two components of the burst history for AXP 1E2259, 
where the solid line is the steep decay (Eq.~\ref{eq:lumin_fast_decay}),
the dash-dotted line is the slow decay (Eq.~\ref{eq:second_comp}), the dotted line is the combination of both
the steep and slow decay, and the dashed line is the quiescent phase luminosity (Eq.~\ref{eq:lxnd}).
The data points are taken from XTE and XMM observations (i.e.~Woods et al. 2004).
}
\end{figure}

 The characteristic timescale for the two components
 in our model in the case of AXP 2259$+$586
  are given by equations (\ref{eq:wallt}) and (\ref{eq:tau13}),
  \begin{eqnarray}
 \nonumber  \tau_{\rm w} &\simeq& 0.11 \quad {\rm hours}\\
                      \tau_{13} &\simeq& 141\quad {\rm days}\ .
  \end{eqnarray}
  The time needed 
to consume 99\% of the wall is  $\sim 100 \tau_{\rm w}\sim 10~\rm{hrs}$, consistent
 with the observed June 2002 outburst. 
 We note that the observed enhanced spin-down of
AXP 1E2259$+$586  lasted for $\sim 18 \pm 6$ days. Naively,
 we expect the enhanced spin-down episode to coincide with the
   wall accretion event. However, the propeller torque
    could in principle act for longer period if there is
    more matter supplied beyond the wall accretion episode.
Shown in Figure~(\ref{fig:2comp}) are the two predicted decay components
 as compared to the observed outburst of AXP 2259$+$586.  The two components in our
 model are given by  equations (\ref{eq:lumin_fast_decay}) and  (\ref{eq:second_comp}).
 
  The second component in our model is 
  consistent with the data, if we adopt $\mu_{\rm atm,b}\sim 2.8$, as shown in Figure~\ref{fig:2comp}.
   Adopting $\mu_{\rm atm,b}\sim 2.8 $  suggests that the   composition of
    the torus atmosphere
    following irradiation  is slightly different from a pure light nuclei
    atmosphere. The   predicted decay of the first component is
slower than the observed one.  We argue that a faster decay rate
could result if one were to include ``squeezing" pressure exerted on the wall
 by the dipole magnetic field. Such an elaborate 
  description of the wall consumption could also lead  to 
 sporadic episodes of  high accretion which could account for  the 
  $\sim 80$ X-ray bursts observed in this source. 
 A minimum timescale for such accretion episodes, corresponding to wall material free-falling
  onto the star along the magnetic field lines, is
\begin{equation}
  t_{\rm acc} =\frac{R_{\rm t}}{v_{\rm ff}}\sim 1\ {\rm ms} \frac{R_{\rm t,15}^{3/2}}{M_{\rm QS,1.4}^{1/2}}
  \sim 2.15\ {\rm ms}\ ,
\end{equation}
 which matches the lower limit in the 
  2 ms to 3 s range in bursts duration observed;  $v_{\rm ff}=\sqrt{2GM/R_{\rm t}}$ is the free-fall velocity.

\section{Discussion and Conclusion}\label{sec:discussion}

With this paper and OLNI, we have completed our investigation of the fate of the ejected crust
during a quark-nova event.  We conclude that, depending on the unknown initial 
conditions of the quark-nova compact remnant 
(specifically the initial spin period, magnetic field strength, and ejecta mass),
the ejected crust (from the parent neutron star) will form either a shell or torus.

Given that the initial spin period that separates torus formation from shell formation
is close to values inferred theoretically (Staff et al. 2006),
one should expect that some SGRs/AXPs possess a shell and some a torus.
In OLNI we concluded that most of the SGRs/AXPs are best modeled by appealing
to the shell picture.  In this paper, we have shown evidence that two
AXPs, 4U0142$+$61 and 1E2259$+$586, are better understood in the torus picture.
These AXPs show X-ray luminosity in excess compared to
what is expected from vortex expulsion, given their rotation period 
and period derivative (see Eq. \ref{eq:lx}).  
We argue that these sources were merely born with initial periods below the 
limiting period given in equation (\ref{eq:plimit}).  

Most of the dynamics and all of the emission mechanisms applicable
to a torus should also apply to the shell case.  However, in
most SGRs/AXPs, all emission is dominated by flux expulsion, so
these effects are unimportant except in AXPs 1E2259+586 and 4U0142+61.
Thus, it may be the case that other SGRs/AXPs also possess a torus.

The recent work by Durant \& van Kerkwijk (2006) indicates that all AXP luminosities
are roughly equal to $1.3\times 10^{35}\rm{erg~s}^{-1}$.  Although these improved luminosity estimates
may indicate a constant luminosity for all AXPs (and possibly the same emission mechanism),
these improved luminosities actually better agree with our theoretical fit (c.f. dotted line in Fig. 1).
Regardless, one still needs to consider the reason for the lower period derivatives of 
AXPs 1E2259+586 and 4U0142+61.
Equation (\ref{eq:lxnd}) predicts that the quiescent accretion luminosity in our model is
independent of the period derivative, so when this
constant accretion luminosity becomes roughly the same as the
spin-down luminosity (eq. \ref{eq:lx}), we cannot distinguish whether
the object has a shell or torus.

Although accretion and propeller models have been considered in the past 
(eg. Mereghetti \& Stella 1995, Mosquera Cuesta et al. 1998, Rothschild et al. 2002), 
as well as models involving exotic compact stars (eg. Usov 2001),
our model is the first to use both in the context of a CFL quark star to explain both bursting and evolution.
Also, there are clear advantages to our model; i) whereas normal accretion disks are generally not 
able to survive the intense gamma bursts, the dense ejecta in our model can not only survive
but also reacts in a manner that provides a good fit to burst profiles.  Also,
ii) the presence of a CFL quark star has the advantage of being able explain both period clustering
and long-term spin-down by appealing to vortex expulsion, as well as possessing the environment
necessary for the type of disk we describe, which accounts for bursting behavior; and
iii) the CFL quark star is extremely volatile, which makes achieving the large burst energies
does not require excessively large magnetic field strengths.

We realize that although it may appear elaborate, the wall-like structure is a result of
the first-order treatment of the interplay between gravity 
and differential rotation. In reality we expect a more complex formation and 
evolution of the torus, but we feel the simplified geometry 
(i.e. ``walls'') helps understand the essentials.
Furthermore, we have not studied the stability of this equilibrium structure.  
We feel that stability could be an issue, however, the model's ability to account 
for observations of AXPs indicates the structure may indeed be stable and similar to what we describe.

A further implication of our model is that the initial
spin period of the quark star is determined by observational parameters.
 Equations (2) and (3) in Niebergal et al. (2006) imply a conserved
 quantity in our model given by
 \begin{equation}
 P B^{2} = P_{\rm 0} B_{\rm 0}^2\ .
 \end{equation}
 Equation above is valid if accretion and propeller torques are small
 compared to magnetic spin-down torques averaged over long timescales.
 Combining this with our equation (\ref{eq:dradius}) yields the quark star birth  period
 \begin{equation}
 P_{\rm i}\simeq 5\ {\rm ms}  \frac{P_{10}\dot{P}_{-13}^{1/2}}{R_{\rm t,15}^{1/4} m_{-7}^{1/2}}\ .
 \end{equation}
  For example in the case of AXP 2259$+$586 for which we have derived the torus
  mass and radius we find a corresponding birth period
  of $\sim 5$ ms for a torus mass of $m_{-7}\sim 2$.  
This is below the limiting period, given by equation (\ref{eq:plimit}), implying torus formation,
  $P_{\rm lim}\sim 8$ ms, providing a self-consistency check of our model. 
Additionally, from the observational point of view, the lack of energetic SNR
surrounding some AXPs and SGRs such as AXP 2259$+$586 (Vink\&Kuiper 2006)
 implies a birth period greater than about 5 ms.

It is not clear if the passive disk discovered around AXP 4U0142 is the torus
we describe here or just a fall-back disk, as in the case of rapidly rotating progenitors,
which could have formed directly from the collapsing SN envelope (before the QN explosion).
However, the low optical emission of the discovered disk, 
is highly suggestive of a degenerate torus as we described in our model.
Further observations of SGRs/AXPs should be able to distinguish between
the disk and the degenerate torus, and would provide an excellent test for our model.

\acknowledgements
This research is supported by grants from the Natural Science and
Engineering Research Council of Canada (NSERC).

\appendix

\section{Torus geometry and evolution}\label{sec:torus_evol}

The formation of the torus involves a vertical expansion at the speed of sound 
until  hydrostatic equilibrium is reached.
For a completely degenerate and relativistic gas the sound speed is
 (see OLNI),
\begin{equation}\label{eqn:soundspeed}
c_{\rm s}^2 =  1.06\times 10^{18} \rho_{10}^{1/3}~{\rm cm^2/s^2}\ ,
\end{equation}
where $\rho_{10}$ is the torus density in units of $10^{10}~\rm{g/cc}$.

While the material expands vertically, momentum conservation 
(i.e. $v_{\rm K}(R_{\rm t}) R_{\rm t} = v_{\rm circ.}(z,r) r$, where $v_{\rm K}$ is the Keplerian velocity) 
causes the material to move 
radially at the same time.  Combined with the circular orbital motion at height $z$, 
\begin{equation}
\label{eq:vcir}
 \frac{v_{\rm circ.}^2}{r}  =  \frac{GM_{\rm QS}}{r^2+z^2}\cos(\theta) \ , 
\end{equation}
where $\cos(\theta) = r/(r^2+z^2)^{1/2}$ and $(r,z)$ are cylindrical coordinates,
the resulting shape of the expanding torus is given by,
\begin{equation}\label{eq:shape}
 \left( 1+\frac{z^2}{r^2}\right)^{3/2} = \frac{r^2}{R_{\rm t}^2}\ .
\end{equation}
From equation (\ref{eq:shape})  above  one can see that the torus flares into a conical shape.
Integrating the equation of hydrostatic equilibrium along the flared surface of the torus
from $(z=0,r=R_{\rm t})$ to $r_{\rm sph.} = \sqrt{r^2+z^2}$ gives the density profile
\begin{equation}\label{eq:rhotorus}
 \rho (r,z)^{1/3}  =   \rho (R_{\rm t},z=0)^{1/3}  - \frac{GM_{\rm QS}}{4 \kappa}  \int_0^{r_{\rm sph}} \frac{\cos(\phi)}{r^2+z^2} ds \ ,
\end{equation}
where $\phi$ is the angle between the tangent to the surface 
(given by Eq.~\ref{eq:shape}) and the radial direction 
($ds$ is the path length element along the surface).

Using equation (\ref{eq:shape}),
the resulting structure extends from $(r,z)\sim (1.4R_{\rm t},-R_{\rm t})$ to $(r,z)\sim (1.4R_{\rm t},+R_{\rm t})$ and
is independent of the torus mass.
In simple terms, the effective  gravity ($g_{\rm eff.} = - GM_{\rm QS}/(r^2+z^2)$)  
is small from $z=0$ to $z\sim R_{\rm t}$ while above $R_{\rm t}$,
where the surface is oriented nearly radially, the effective gravity is strong resulting 
in a rapid decline in density.  
The timescale for vertical expansion is 
$R_{\rm t}/c_{\rm s}\sim (1.5\ {\rm ms})\times R_{\rm t,15}/\rho_{t, 10}^{1/6}$ where
$\rho_{\rm t,10}$ is the torus density in units of $10^{10}$  g cm$^{-3}$..

 The radial thickness of the torus, $\Delta r_{\rm, t}$,  is
governed by the angular momentum transfer due to viscosity. 
The viscosity due to particle collisions in a degenerate gas is estimated using that for 
an ideal gas (e.g.~Lang pg.~266)\footnote{We note that turbulent
viscosity is negligible in our case since the torus is very dense and metal rich. The 
corresponding $\alpha$ parameter (Shakura \& Sunyaev 1973) is very small and is given as 
$\alpha = \nu/c_{\rm s}/H = 6\times 10^{-4} T_{\rm MeV}^2/H_{10}$ where
the disk scale height $H$ is in unist of 10 km.},
\begin{equation}\label{eq:nu}
  \nu=  6\times 10^{11} T_{\rm MeV}^{5/2}\ {\rm cm}^{2}\ {\rm s}^{-1}\  .
\end{equation}
 Equation (\ref{eq:nu}) shows the strong dependence of the viscosity on the temperature, thus
 on cooling. The radial expansion of the torus  can be derived using 
$d\left( (\Delta r)^2 \right)/dt = \nu$.  Due to the short timescale  associated with the
 propeller phase, (eq. \ref{eq:tprop}), the radial width prior to vertical expansion is
\begin{equation}\label{eq:drtorus}
   (\Delta r)_{\rm prop} \simeq (\nu t_{\rm prop})^{1/2}\sim   1.7\times 10^{4} \ {\rm cm}  \frac{T_{\rm MeV}^{5/4} R_{\rm t,15}^{3/8}}{M_{\rm QS,1.4}^{1/8}}\ .
\end{equation}
As shown in OLNI, the shell temperature at formation is of the order
 of 1 MeV. Blackbody cooling of the shell during the propeller phase yields a temperature
 of the propelled matter decreasing with time from an initial $\sim 1$ MeV to
 a final $\sim 60$ keV. Including cooling during the propeller phase slightly
 affects $(\Delta r)_{\rm prop.}$; the constant in front of equation (\ref{eq:drtorus})
 reduced to $1.2\times 10^{4}$ cm. This is understandable since the torus
 spreading is dominated by the early hot part of the propeller phase. 
 During the vertical expansion, with an initial temperature
 of $\sim 60$ keV, viscous spreading results in an additional
  $(\Delta r)_{\rm exp}\sim 10^{3}$ cm.

Following the vertical expansion, accretion from the edges of the torus begins. 
During this time the torus is also cooling as a blackbody,
 and  quickly reaches an equilibrium temperature, $T_{\rm eq}$, in the
 keV range as given in Appendix A
  (eq.(\ref{eq:Teq})).
Subsequent viscous spreading is then determined by $T_{\rm eq}$.  
Thus the spreading for some equilibrium temperature is,
\begin{equation} \label{eq:drkev}
  (\Delta r)_{\rm t}\sim  7.8\times 10^{5}\ {\rm cm} \ T_{\rm keV}^{5/4} t_{\rm yrs}^{1/2}\ ,
\end{equation}
where $t_{\rm yrs}$ is the age of the system in years\footnote{We recall that in our model the age of the system is given by  
$ t_{\rm age} \sim 10^{6}\ {\rm yrs}\ \frac{P_{10}}{\dot{P}_{-13}}$ for $t>> \tau$
 where $\tau$ is the magnetic field decay timescale given in Niebergal et al. (2006).
  Here, the period and the spin-down rate are units of 10 seconds and
 $10^{-13}$ s/s, respectively.}, and $T_{\rm keV}$ is the equilibrium temperature
of the torus in keV. 
The torus average density decreases in time as, 
\begin{equation}\label{eq:rhot}
 \rho = \frac{m}{4\pi R_{\rm t}^2 \Delta r_{\rm t}} 
 \simeq 9.1\times 10^{6}\ {\rm g\ cm}^{-3} \frac{m_{\rm -7}}{R_{\rm t,15}^2 T_{\rm keV}^{5/4}t_{\rm yrs}^{1/2}}\ .
\end{equation}

In general at a given temperature the maximum density, $\rho_{\rm nd, q}$, of the torus below
which the matter is  non-degenerate is found by setting  $T_{\rm t} = T_{\rm Fermi}
 = 123.6\ {\rm MeV} \mu_{\rm e}^{-2/3} \rho_{10}^{2/3}$ (see appendix A), or,
\begin{equation}\label{eq:rhondtext}
  \rho_{\rm nd}  \simeq  230 \mu_{\rm e}\ {\rm g\ cm}^{-3} T_{\rm keV}^{3/2}\ ,
\end{equation}
where $\mu_{\rm e}=2$ is the mean mass per electron.
Since the torus equilibrium  temperature stays within the
keV range it would take $\sim 10^{9}$ years to become non-degenerate,
ensuring that the cold torus always remains in a degenerate state.

\section{Torus equilibrium temperature}

In general at a given temperature the maximum torus density below
which the matter is  non-degenerate, $\rho_{\rm nd}$, is found by equating the torus temperature 
to the Fermi temperature, 
$T = T_{\rm Fermi} = 123.6\ {\rm MeV} \mu_{\rm e}^{-2/3} \rho_{10}^{2/3}$ (see appendix in OLNI), or,
\begin{equation}\label{eq:rhond}
  \rho_{\rm nd}  \simeq  460\mu_{\rm e} \ {\rm g\ cm}^{-3} T_{\rm keV}^{3/2} \ .
\end{equation}
The corresponding scale height of the torus atmosphere is
\begin{equation}\label{eq:hnd}
H_{\rm atm} = \frac{v_{\rm th.}^{2}}{g_{\rm t}}\sim 12\ {\rm cm}\ 
\frac{T_{\rm keV}R_{\rm t,15}^{2}}{\mu_{\rm atm} M_{\rm QS,1.4}}\ ,
\end{equation}
where $g_{\rm t} = GM_{\rm QS}/(\sqrt{2}R_{\rm t})^2\sim 4\times 10^{13} M_{\rm QS.1.4}/R_{\rm t,15}^2\ {\rm cm~s}^{-2}$
 is the effective gravity at the torus, and
 \begin{equation}
v_{\rm th}=\left(\frac{kT}{\mu_{\rm atm} m_{\rm H}}\right)^{1/2} \sim 
\frac{3.1\times 10^{7}}{\mu_{\rm atm}^{1/2}}\ {\rm cm\ s}^{-1}
T_{\rm keV}^{1/2}\ .
\end{equation}
The corresponding column density of iron atoms in the non-degenrate atmosphere is
\begin{equation}
N_{\rm Fe} = \rho_{\rm nd} H_{\rm atm} \sim 
5\times 10^{25}\ {\rm cm}^{-2}\ \frac{T_{\rm keV}^{5/2}R_{\rm t,15}^{2}}{\mu_{\rm atm} M_{\rm QS,1.4}}\ .
\end{equation}

\subsection{Atmosphere opacity}

The torus atmosphere is highly optically thick to radiation  ($\tau_{\rm atm} \ge 10^{7}$
 for temperatures of order 1 keV, where $\tau_{\rm m}$ is the  total optical depth of the atmosphere;
 e.g. \S 3.3 in Clayton 1983).   For conduction, the opacity is lower thus conduction determines  the
  atmospheric temperature profile. For densities $\rho \sim \rho_{\rm nd}$ and temperatures
   in the keV range relevant here, the conductive opacity
   is found to be of the order of $\kappa_{\rm c} \sim 200\ {\rm cm^{2}\ gm}^{-1}$ (see \S 3.4 in Clayton 1983).
    The resulting conductive optical depth is $\tau_{\rm c}= \kappa_{\rm c} \rho_{\rm nd}H_{\rm t,
    atm.}\sim 6\times 10^{5}$. The standard gray atmosphere model
 gives $T_{\rm atm}\propto \tau^{1/4}$ where $\tau$ is the optical depth in the atmosphere. 
 This ensures  the effective temperature at the top of the atmosphere  to be less than $1/10$ of that of the bulk body of the torus. Since the blackbody cooling rate goes as $T^4$, the cooling is dominated by
 cooling from the sides of the torus until the area of the atmosphere greatly exceed
 that of the sides of the torus, or, $\Delta r_{\rm t} > 10^{4} R_{\rm t}$.  Using equation (\ref{eq:drkev})
  this occurs at a time, in years, $t> 3.6\times 10^{8}\ {\rm yrs}\ \times R_{\rm t,15}^{2}/T_{\rm keV}^{5/2}$,
 much later than the time relevant here.
 
 \subsection{Torus thermal evolution}
 
The torus cooling is given by,
\begin{equation}\label{eq:lbb}
L_{\rm BB,t} = 4\pi R_{\rm t}^{2}\sigma T^4\sim  2.9\times 10^{37}\ {\rm erg\ s}^{-1}~R_{\rm t,15}^{2} T_{\rm keV}^{4}\ .
\end{equation}

We will assume that the heated torus atmosphere leaks out from the sides
of height $H_{\rm nd}$. The accretion rate is then 
\begin{equation}\label{eq:lxacc}
\dot{m} = 2\pi R_{\rm t} 2 H_{\rm nd} \rho_{\rm nd} v_{\rm th}
 \sim  1.7\times 10^{17}\ {\rm g\ s}^{-1}\  \frac{R_{\rm t, 15}^{3} T_{\rm keV}^{3}}{\mu_{\rm atm}^{3/2}M_{\rm QS,1.4}}\ ,
 \end{equation}
resulting on an accretion luminosity of
\begin{equation}
L_{\rm acc} =  \eta \dot{m} c^2\sim 1.5\times 10^{37}\ {\rm erg\ s}^{-1}  
\frac{\eta_{0.1} R_{\rm t, 15}^{3} T_{\rm keV}^{3}}{\mu_{\rm atm}^{3/2}M_{\rm QS,1.4}}
\end{equation}
where we estimate that 10\% (i.e. $\eta =0.1$) 
of the mass-energy of accreted matter is released as radiation.

The thermal evolution of the torus is given by the heat capacity of the torus,
\begin{equation}\label{eq:heat}
C_{\rm v}  \frac{\partial T}{\partial t}  = L_{\rm acc} - L_{\rm BB,t}\ .
\end{equation}
In the equation above $C_{\rm v} = N_{\rm t} c_{\rm v}$ is the torus heat capacity
with  $N_{\rm t}= m/(\mu_{\rm e} m_{\rm H})$ and it's
 specific heat is $c_{\rm v}= (k_{\rm B}T/\epsilon_{\rm F})\times k_{\rm B}\pi^2/2$, 
 where $k_{\rm B}$ is the Boltzmann constant and $\epsilon_{\rm F}$ the Fermi energy.
  The heat capacity of the torus is then
\begin{equation}\label{eq:cv}
C_{\rm v}  \sim 5.2\times 10^{31}\ {\rm erg\ K}^{-1}\frac{T_{\rm keV} m_{\rm -7}}{\rho_{\rm 7}^{2/3}}\ .
\end{equation}
Applying equation (\ref{eq:rhot}) gives,
\begin{equation}\label{eq:cv2}
C_{\rm v,t}  \sim 5.5\times 10^{31}\ {\rm erg\ K}^{-1}\ T_{\rm keV}^{11/6} 
m_{\rm -7}^{1/3} R_{\rm t,15}^{4/3} t_{\rm yrs}^{1/3}\ .
\end{equation}
Thus, equation (\ref{eq:heat}) becomes,
\begin{equation}\label{eq:heat2}
\frac{\partial T_{\rm keV}}{\partial t^{2/3}} = A T_{\rm keV}^{7/6} - B T_{\rm keV}^{13/6}\ ,
\end{equation}
where 
\begin{eqnarray}
A &=&  7.43 \frac{\eta_{0.1}R_{\rm t,15}^{5/3}}{\mu_{\rm atm}^{3/2}
m_{\rm -7}^{1/3}M_{\rm QS,1.4}} \\\nonumber
B &=&  14.4 \frac{R_{\rm t,15}^{2/3}}{m_{\rm -7}^{1/3}}\ .
\end{eqnarray}
 The equilibrium temperature can be found by setting $\partial/\partial t=0$ to get
\begin{equation}\label{eq:Teq}
T_{\rm eq} =\frac{A}{B} \sim  0.52 \ {\rm keV} \frac{\eta_{0.1}R_{t,15}}{\mu_{\rm atm}^{3/2}M_{\rm QS,1.4}}\ .
\end{equation}
  The torus reaches the equilibrium temperature almost immediately after accretion ensues
  in timescales of the order of $\sim 1/A$.



\section{Wall equilibrium temperature}

Once the wall is penetrated by the magnetic field and detached from the torus
it will co-rotate with the star. Unlike in the torus itself, where the atmosphere is removed from
the sides, in the case of the wall the atmosphere evaporates directly along the magnetic field lines from the ends.
The mass-loss rate of the wall through its atmosphere can then be written as,
 \begin{equation}\label{eq:mdotwall}
\dot{m}_{\rm w} = 2\pi R_{\rm t} \Delta r_{\rm w} \rho_{\rm w, nd} v_{\rm w,th} = 6.85\times 10^{18}~\rm{g/cc}~
\frac{R_{\rm t,15}^{7/2} T_{\rm keV}^2 m_{\rm w,-10}}{\mu_{\rm w,atm}^{1/2} B_{\rm s,14}^{3/2} R_{\rm QS}^{9/2} } \ .
 \end{equation}
Here, $\rho_{\rm w,nd} = 460~\rm{g/cc}~T_{\rm w, keV}^{3/2}$ (from Eq.~(\ref{eq:rhond})),
$v_{\rm w,th}^2 = kT_{\rm w}/(\mu_{\rm w,atm.}m_{\rm H})$ is the thermal velocity,
and $\Delta r_{\rm w} = 2\pi R_{\rm t}^2 m_{\rm w}/\rho_{\rm w}$ due to the wall becoming thinner as it loses mass while
maintaining a constant density (as described in \S~\ref{sec:wallaccretion}).  

The resulting accretion luminosity $L_{\rm w}=\eta \dot{m}_{\rm w}c^2$  heats up the wall
which is meanwhile cooling as a blackbody at a rate of $L_{\rm w, BB}
   \sim 2.71\times 10^{37}\ {\rm erg\ s}^{-1} R_{\rm t,15}^2 T_{\rm keV}^{4}$ (we assume
    the wall to cool mostly from its side facing the star).
The wall's equilibrium temperature is then obtained by setting $L_{\rm w}= L_{\rm w,BB}$ which yields
\begin{equation}\label{eq:Twall}
T_{\rm w,eq} \sim 4.74\ {\rm keV}\ 
\frac{\eta_{0.1}^{1/2} R_{\rm t,15}^{3/4}m_{\rm w,-10}^{1/2}}
{\mu_{\rm w,atm}^{1/4}B_{\rm s,14}^{3/4}R_{\rm QS,10}^{9/4}} \ .
\end{equation}

\end{document}